\documentclass[3p,times,twocolumn]{elsarticle}

\usepackage{ecrc-ac_arxiv,widetext}
\usepackage{mathrsfs,latexsym}
\usepackage{booktabs}
\newcommand{\bew}{\begin{widetext}\begin{equation}}
\newcommand{\eew}{\end{equation}\end{widetext}}

\volume{00}

\firstpage{1}

\journalname{Nuclear Physics B Proceedings Supplement}

\runauth{M. Della Morte, J. Heitger, H. Simma and R. Sommer}

\jid{nuphbp}

\jnltitlelogo{Nuclear Physics B Proceedings Supplement}

\usepackage{amssymb,bbm}
\usepackage{amsbsy,color}

\usepackage{macros_hs}

\begin{document}

\begin{frontmatter}



\dochead{\small{\flushright{SFB/CPP-14-102\\ 
DESY 15-007 \\ CP3-Origins-2014-048 DNRF90\\DIAS-2014-48\\MS-TP-15-01\\}}}


\title{
Non-perturbative Heavy Quark Effective Theory:
\\
An application to semi-leptonic B-decays
}


\author[DK,ES]{Michele Della Morte}
\author[MU]{Jochen Heitger}
\author[DE]{Hubert Simma}
\author[DE,HU]{Rainer Sommer}

\address[DK]{CP3-Origins \& Danish IAS, University of Southern Denmark, Campusvej 55, DK-5230 Odense M, Denmark}
\address[ES]{IFIC (CSIC), c/ Catedr\'atico Jos\'e Beltr\'an, 2. E-46980, Paterna, Spain}
\address[MU]{Westf\"alische Wilhelms-Universit\"at M\"unster, Institut f\"ur Theoretische Physik,
Wilhelm-Klemm-Stra{\ss}e 9, D-48149 M\"unster, Germany}
\address[DE]{John von Neumann Institute for Computing (NIC), DESY, Platanenallee~6, D-15738 Zeuthen, Germany}
\address[HU]{Institut f{\"u}r Physik, Humboldt Universit{\"a}t, 
Newtonstra{\ss}e 15, D-12489 Berlin, Germany}

\begin{abstract}
We review a lattice strategy how to non-perturbatively determine the coefficients
in the HQET expansion of all components of the heavy-light
axial and vector currents, including $\mhinv$-corrections.
We also discuss recent preliminary results on the form factors parameterizing semi-leptonic B-decays at the leading order in $\mhinv$. 

\end{abstract}


\end{frontmatter}


\section{Introduction}
\label{intro}

B-meson decays present an excellent opportunity to 
test the Standard Model and to search for New Physics. 
Many decay channels are allowed.  Their rates depend on  
CKM matrix elements and often have a sensitivity to heavy 
particles both within and outside the Standard Model. 
Such effects provide us with indirect tests of New Physics
which are complementary to direct 
searches for new particles at very high energy. In this way,
B-decays provide for stringent
constraints on the structure and couplings of theories beyond the Standard Model.
Of course, to make this program work, the theory of these decays
needs to be understood. Here the challenge is in the correct 
handling of QCD.

The relevant intrinsic QCD scale is much smaller than the mass scale of the electroweak 
interactions or any New Physics. Therefore, the effect of beyond the Standard Model interactions can accurately be described by a perturbatively 
treated effective Hamiltonian,
consisting, e.g., of bilinear quark currents 
or local four-quark interactions. 
One then has to compute the hadronic matrix elements of
this effective Hamiltonian with the basic theory given
just by QCD (with 
$N_{\rm f}=5$ flavours).

A first principles evaluation of the matrix elements is 
possible by lattice QCD, non-perturbatively in the QCD coupling 
but, as mentioned, perturbatively in the effective Hamiltonian,
which has contributions from the electroweak sector and, possibly,
from New Physics.
We do, however, still have to treat a hierarchy of scales
\bd
  L^{-1}  \ll m_\pi \approx 140\,\mathrm{MeV} \ll m_\mathrm B \approx 5\,\mathrm{GeV} \ll a^{-1} \ .
\ed
This forbids direct simulations on a lattice with spatial extent $L$ 
and lattice spacing $a$ on nowadays computers. We therefore apply 
another effective theory, HQET, which provides us (at appropriate kinematics) with
the expansion of QCD observables in inverse powers of $\mh$. The mass scale
in the numerator of this expansion parameter is the intrinsic QCD-scale of
around $500\,\MeV$ as well as (in the rest frame of the B-meson) small
spatial momenta. This theory, put on the lattice, only has to 
cover accurately the scales far below $m_\mathrm B$.

HQET has low-energy constants (or HQET parameters) which need to and can be determined 
non-perturbatively, as explained in \cite{sfb:C1a}. 
Besides the HQET parameters in the action, there are a number of parameters
for each considered composite field which makes up the
electroweak Hamiltonian.
Extending the procedure discussed in \cite{sfb:C1a}
for the parameters in the action and the temporal component of the axial
current, we review here a matching strategy for the full set of 
temporal and spatial components of the heavy-light vector and axial currents.

These are of relevance for
the determination of the CKM elements. For instance, $V_{\rm ub}$
can be determined from the inclusive decays 
$\mathrm B\rightarrow X_\mathrm u \ell \nu$ \cite[sect. 12]{Agashe:2014kda} 
\be
  |V_\mathrm{ub}|= (4.41 \pm0.15^{+0.15}_{-0.19})\times 10^{-3} \ ,
  \label{VubPDGincl}
\ee
or from various exclusive decays. From the exclusive
semi-leptonic decay $\mathrm B \rightarrow \pi \ell \nu$,
one finds \cite[sect. 12]{Agashe:2014kda} 
\be
  |V_\mathrm{ub}|= (3.28 \pm 0.29)\times 10^{-3} \ ,
  \label{VubPDGexcl}
\ee
while the purely leptonic channel, which is more difficult in experiment,
tends to give values of $\vert V_{\rm ub}\vert$ that are significantly larger
than in eq.~(\ref{VubPDGexcl}).
For instance, using the branching ratio 
${\cal B}(\mathrm B\to\tau\nu) = (1.14 \pm 0.22)\times 10^{-4}$
one obtains \cite[sect. 12]{Agashe:2014kda} 
\be
    |V_\mathrm{ub}|= (4.22 \pm 0.42)\times 10^{-3} \ . 
\ee
The discrepancies between these different determinations
of $|V_\mathrm{ub}|$ are not understood. We need both theoretical and experimental 
improvements to find out their origin. For exclusive decays, such as
$\mathrm B \rightarrow \tau \nu$ or $\mathrm B \rightarrow \pi \ell \nu$,
it is known how to compute the required hadronic matrix elements 
by lattice computations. In this respect, the purely leptonic decay channel 
$\mathrm B \rightarrow \tau \nu$, is the simplest case and
the required decay constant $f_{\mathrm B}$ is known precisely.
We refer to \cite{Aoki:2013ldr} and references therein for a summay.

Semi-leptonic decays, such as $\mathrm B \rightarrow \pi \ell \nu$ or 
$\mathrm B_s \rightarrow \mathrm K \ell \nu$ allow to extract 
the CKM parameter $|V_\mathrm{ub}|$ from a combination of the 
experimental differential decay 
rate with a theoretical determination of the form factor $f_+(q^2)$. 
In particular lattice QCD can achieve the computation of $f_+(q^2)$.
In principle, it is sufficient to do this at a single value of $q^2$, the squared 
momentum transfer. However,
in practice, experimental data is provided over a 
range (of bins) of $q^2$, and one has to use some knowledge on 
the functional form of  $f_+(q^2)$, e.g., in the form of the
BCL paramterisation \cite{Bourrely:2008za}. Then, a 
theoretical prediction of $f_+(q^2)$ for a single $q^2$ is sufficient to extract 
$|V_\mathrm{ub}|$. In section 4, we investigate the feasibility of a precise lattice 
determination of the form factor in the continuum limit and at a fixed $q^2$. 
Of course, eventually the precision will be enhanced when the form factor is known 
at more than one $q^2$.

\newcommand{\nspace}{\hspace*{-5mm}}

\section{General matching strategy}
\label{gen-strat}

Following~\cite{Heitger:2003nj,Blossier:2010jk,Blossier:2012qu,DellaMorte:2013ega}, 
we fix the 19 parameters in the HQET expansion of the action and vector/axial currents
through matching equations, which we write in the form
\be
\Phi^\qcd_i(L,m_{\rm h},0)=
\Phi^\hqet_i(L,m_{\rm h},a) \,.
\label{e:matchingcond}
\ee
In the above equations the $\Phi_i\; (i=1\dots 19)$ are finite-volume, renormalized, QCD quantities defined 
in the continuum
\be
      \Phi^\qcd_i(L,m_{\rm h},0) = \lim_{a\to 0}\Phi^\qcd_i(L,m_{\rm h},a)\;
\ee
whereas the  $\Phi^\hqet_i$ are understood to be expanded
up to a given order in $1/m_{\rm h}$ (NLO in our setup) and computed in HQET
at a finite lattice spacing. The $\mhinv$-expansion of the observables $\Phi_i$ results
from expanding the action and composite fields $O^\qcd(x)$, viz.,
\bea
\Oop^\hqet(m_{\rm h})&=& 
Z_O\left\{\Oop^\stat+\sum_n c_n\Oop_n\right\} \,.
\label{e:ohqet}
\eea
The coefficients in the expansion are chosen such that 
\eq{e:matchingcond} holds. 
In the r.h.s.~of the above equation one has to include all linearly independent 
operators with mass dimension one higher than $O^\qcd$ (or $\Oop^\stat$) 
transforming in the same way as $O^\qcd$ under the common set of 
symmetries of QCD and HQET.
These are needed for the renormalization and $\Order{a}$ improvement 
of the effective theory, and in order to account 
in HQET for the $m_{\rm h}$-effects of QCD. Such effects are described by
the $m_{\rm h}$-dependence of the parameters $Z_O$ and $c_n$
in eq.~(\ref{e:ohqet}).
Once the matching is performed at a given value of the heavy quark mass 
and fixed $L=L_1$, the $L$-dependence in the difference
$\Phi^\qcd_i(L,m_{\rm h},0)-\Phi^\hqet_i(L,m_{\rm h},a)$ quantifies higher-order (in $\mhinv$) corrections.  
These have been studied at tree-level in~\cite{DellaMorte:2013ega} for a particular class of matching conditions. Note that at tree-level and in the continuum
limit, the $L$-dependence of dimensionless observables $\Phi_i$ appears
only in the combination $z=L m_{\rm h}$.
\subsection{Definition of the matching observables}
Let us assume we want to determine the parameters appearing at NLO in the expansion
of a current $\Jop$.

The matching observables can be constructed from suitable combinations 
of correlation functions, $\mathcal C_J$, which typically have a single
insertion of $\Jop$. Writing the HQET expansion of $\Jop$ as
\be
    J^\hqet = 
    Z^\hqet_J \left\{ \Jop^\stat + \sum_n c_{J_n} \Jop_n \right\}  
    + \Order{1/m_{\rm h}^2}\,,
    \label{e:jexpansion}
\ee
then the correlation functions take the generic form
\bea
   && \mathcal C^\hqet_J =  
      Z^\hqet_J Z^\hqet_{\mathcal C} e^{-m_\bare x_{\mathcal C}} 
      \big\{ \mathcal C^\stat_J + \nonumber \\
   && \omegakin \mathcal C^\kin_J + \omegaspin \mathcal C^\spin_J 
      + \sum_n c_{J_n} \mathcal C^\nlo_{J_n} \big\}\,,
      \label{e:cfexpansion}
\eea
where all correlators on the r.h.s. are computed in the static
approximation.
The only appearance of the bare HQET quark mass, $m_\bare$, 
is in the factor $e^{-m_\bare x_{\mathcal C}}$, 
with $x_{\mathcal C}$ given by the time distances of the heavy (static) quark fields 
entering in $\mathcal C^\hqet_J$.
Apart from 
$Z^\hqet_J$, all other (re)normalization factors contributing
to $\mathcal C^\hqet_J$ are collected in $Z^\hqet_{\mathcal C}$.

In the correlation functions on the r.h.s. of eq.~(\ref{e:cfexpansion}),
the leading-order term $\mathcal C^\stat_J$ has just one insertion of $\Jop^\stat$
(instead of $\Jop$), while $\mathcal C^{\kin/\spin}_J$ differ from $\mathcal C^\stat_J$ by an 
extra insertion (summed over the entire space-time volume) of the $1/m_{\rm h}$-terms 
$\Okin$ or $\Ospin$ from the Lagrangian. 
The other next-to-leading contributions $\mathcal C^\nlo_{J_n}$ have an insertion of 
one of the higher-dimensional operators $\Jop_n$
from the expansion in eq.~(\ref{e:jexpansion}). 

The observables $\Phi_i$ are  defined as combinations of such correlators, typically logarithms of ratios,
and only  one of the observables, say $\Phi_1$, is left with an explicit dependence on $m_\bare$.

By combining all HQET parameters into a vector 
\be
   \!\!\!\!\!\!\!\!\!\!\omega = (m_\bare,\ \omegakin,\ \omegaspin,\ c_{J_1},\ \ldots,\ \ln Z^{\rm HQET}_{J},\ \ldots )^T,
\ee
the HQET expansion of the observables can be cast in the form
\bea
 &&\Phi_i^\hqet(L,M,a) = \eta_i(L,a) + \nonumber \\
&&\varphi_i^j(L,a)\,\omega_j(M,a) + \Order{1/m_{\rm h}^2}\,, 
 \label{e:phiexp}
\eea
where $M$ is the Renormalization Group Invariant (RGI) heavy quark mass and
the vector $\eta$ represents the contribution of the static terms 
$\mathcal C_J^{\rm stat}$ in the correlators involved.
In general, the existence of a continuum limit can be guaranteed only for the combination of HQET 
quantities which appears on the r.h.s. of eq.~(\ref{e:phiexp}).
Only at tree-level
each individual term on the r.h.s. has a well defined continuum limit.

In order to learn about the structure of the matrix $\varphi_i^j$ in eq.~(\ref{e:phiexp}), we group the HQET parameters into
blocks $(m_\mrm{bare})$, $(\omegakin,\,\omegaspin)$,
 $(c_{J_{i}},\,Z_J )$, $(c_{J^\prime_{i}},\,Z_{J^\prime} )$,
$(c_{J^{\prime\prime}_{i}},\,Z_{J^{\prime\prime}} )$, \ldots,
and assume that $J$ is the current which is used in $\Phi_1$.
A suitable choice of the other observables then yields the following  form of
the matrix $\varphi_i^j$:
\bi
\item All entries of the first column, except  $\varphi_1^1$, vanish.
\item In the first row, $\varphi_1^2$ and $\varphi_1^3$ are non-vanishing, due to 
      the contributions from $\omegakin$ and $\omegaspin$ to $\Phi_1$.
      In addition, there may be non-zero $\varphi_1^j$ with $j$ from a single
      block, which corresponds to the current used in $\Phi_1$.
      In our case, this comes from the $A_{0,1}$-term
      (i.e., a term at NLO in the HQET expansion of the temporal component
      of the axial current, analogous to eq.~(\ref{e:v01}) below), 
      which enters the matching condition for $m_\mrm{bare}$.
\item The rest has a simple  block structure, with non-zero blocks
      only in the second block-column (corresponding to contributions from
      $\omegakin$ and $\omegaspin$) and in the blocks on the diagonal 
      (corresponding to the mixings in the last term of 
      eq.~(\ref{e:cfexpansion})).
\ei
Schematically, we
have the following block structure:
\begin{center}
\begin{displaymath}
  \varphi = \left(
     \begin{array}{c|c|c|c|c}
         \varphi_1^1 & \ast & \ast  &   0   &    0    \\ \hline
         0            & \ast &   0   &   0   &   0    \\ \hline
         0            & \ast & \ast  &   0   &   0    \\ \hline
         0            & \ast &   0   & \ast  &   0    \\ \hline
         0            & \ast &   0   &   0   & \ast   \\
     \end{array}
  \right)\;.
\end{displaymath}
\end{center}
That form can be preserved when an
additional (effective) operator $\Jop^\prime$ is included in the matching.
Therefore, the  system can always be solved, and the HQET parameters can be determined, 
by block-wise backward substitution.

Let us explicitly discuss the case of the (renormalized) heavy-light 
vector current in HQET. The temporal component is
\bea
  \label{e:v0hqet}
  \!\!\!\!\! V^\hqet_0(x) = Z_{{\rm V}_0}^\hqet\left[\,V^\stat_0(x)+\sum_{i=1}^2\ceff{V}{0}{i} V_{0,i}(x)\,\right] \,,
\eea
with the static term
\bea
 V^\stat_\mu(x)  & = & \psibarlight(x)\gamma_\mu\psiheavy(x)
\eea
and two additional dimension-four contributions
\bea
\label{e:v01}
 V_{0,1}(x) &=& \psibarlight(x)\frac{1}{2}
            \gamma_i(\nabsym{i}-\lnabsym{i})\psiheavy(x)\,,
\\
 V_{0,2}(x) &=& \psibarlight(x)\frac{1}{2}
            \gamma_i(\nabsym{i}+\lnabsym{i})\psiheavy(x)\,,
\eea
where all derivatives are defined from symmetric nearest neighbour 
differences and $\psiheavy$ and $\psilight$ denote static and light 
(relativistic) quark fields, respectively. 
In order to have access to a variety of different kinematical configurations, 
we will make extensive use here of
generalized periodic boundary conditions for the fermions
\be
\!\!\!\!\psi(x+L\hat{k})=e^{i\theta_k}\psi(x)\,, \quad \overline{\psi}(x+L\hat{k})=\overline{\psi}(x)e^{-i\theta_k}.
\label{e:theta}
\ee
The phases $\theta_k$ \cite{pert:1loop,Luscher:1996sc}, sometimes referred to as {\em twisting}~\cite{Bedaque:2004kc,Sachrajda:2004mi},
 can be employed to inject a momentum $|\Th -\Tq|/L$
in correlation functions~\cite{Bedaque:2004kc,Sachrajda:2004mi}.
For example, it is clear that for $\Th=\Tq$ composite fields such as $V_{0,2}(x)$ above can be associated with total 
derivative operators and, therefore, correlation functions at zero external momenta are not sensitive to their insertions.

Considering now the spatial components of the vector current, the
HQET expansions (see~\cite{Sommer:2010ic}) can be written as
\be
\!\!\!  V^\hqet_k(x) = Z_{\vec{\rm V}}^\hqet\left[\,V_k^\stat(x)+ \sum_{i=1}^4 c_{{\rm V}_{k,i}} V_{k,i}(x)\,\right]\,, 
  \label{e:vkhqet}
\ee
with the following four extra terms:
\bea
\nonumber
 V_{k,1}(x) & = & \psibarlight(x){1\over2}
            (\nabsym{i}-\lnabsym{i}) \gamma_i\gamma_k  \psiheavy(x)\,,
\\
\nonumber
 V_{k,2}(x) & = & \psibarlight(x){1\over2}
            (\nabsym{k}-\lnabsym{k})\psiheavy(x)\,,
\\
\nonumber
 V_{k,3}(x) & = & \psibarlight(x){1\over2}
            (\nabsym{i}+\lnabsym{i}) \gamma_i\gamma_k  \psiheavy(x)\,,
\\
\nonumber
V_{k,4}(x) & = & \psibarlight(x){1\over2}
            (\nabsym{k}+\lnabsym{k})\psiheavy(x)\,.
\eea
The axial vector current components are listed in \cite{DellaMorte:2013ega}.
The classical values of the coefficients are
\bea
&&\ceff{V}{0}{1}=\ceff{V}{0}{2}=\ceff{V}{k}{1}=\ceff{V}{k}{3}
={1 \over 2m_{\rm h}} \,, \nonumber \\
&&{\rm and} \quad 
\ceff{V}{k}{2}=\ceff{V}{k}{4}= - {1 \over m_{\rm h}} \;.
\eea

Concretely, as in~\cite{Heitger:2003nj,Blossier:2010jk,Blossier:2012qu}, 
it is advantageous to construct the matching observables using 
Schr\"odinger functional (SF) homogeneous boundary conditions at $x_0=0$ and $x_0=T$
\cite{Luscher:1992an,Sint:1993un,Sint:1995rb}.
Two categories of QCD correlation functions are formed from composite fields in the bulk, $0<x_0<T$, and boundary quark fields $\zeta \ldots \zetabarprime$.
\begin{itemize}
\item Boundary-to-boundary correlators
\bes
\hspace*{-10mm}
\Fone(\T_\ell,\thb)  = \hspace*{12mm}
\nonumber &&  \nonumber \\ && \nonumber \hspace*{-35mm}
  -{a^{12} \over 2L^6}\sum_{\vecu,\vecv,\vecy,\vecz}
     \left\langle
     \zetabarprime_\light(\vecu)\gamma_5\zetaprime_{\rm b}(\vecv)\,
     \zetabar_{\rm b}(\vecy)\gamma_5\zeta_\light(\vecz)
     \right\rangle\,, 
\nonumber \\
\hspace*{-10mm}
   F^{\light\light}_1({\T}_{\ell},{\T}_{\ell'}) = \hspace*{10mm}
\nonumber &&  \nonumber \\ && \nonumber \hspace*{-35mm}
     -{a^{12} \over 2L^6}\sum_{\vecu,\vecv,\vecy,\vecz}
     \left\langle
     \zetabarprime_\light(\vecu)\gamma_5\zetaprime_{\ell'}(\vecv)\,
     \zetabar_{\ell'}(\vecy)\gamma_5\zeta_\light(\vecz)
     \right\rangle\,, 
\nonumber \\
\hspace*{-10mm}
\Kone(\T_\ell,\thb) = \hspace*{12mm}
\nonumber &&  \nonumber \\ && \nonumber \hspace*{-35mm}
    -{a^{12} \over 6L^6}\sum_{k}\sum_{\vecu,\vecv,\vecy,\vecz}
     \left\langle
     \zetabarprime_\light(\vecu)\gamma_k\zetaprime_{\rm b}(\vecv)\,
     \zetabar_{\rm b}(\vecy)\gamma_k\zeta_\light(\vecz)
     \right\rangle\,, 
\nonumber \\
  \hspace*{-5mm}  \f1v0(x_0,\T_{\ell},\T_{\ell'},\thh) =
\nonumber && \nonumber \\ && \nonumber \hspace*{-35mm} 
 -{{a^{15}}\over{2L^6}}\sum_{\vecu,\vecv,\vecy,\vecz,\vecx} 
  \left\langle \zetabar'_{\ell'}({\vecu}) \gamma_5 \zeta'_{\ell} ({\vecv})
  (V_{\rm I})_0(x) \zetabar_{\rm b} ({\vecy})\gamma_5 \zeta_{\ell'}({\vecz})
  \right\rangle \,,
\label{e_f1v0}
\ees
\item Bulk-to-boundary correlators
\bes
\hspace*{-5mm}\kv0(x_0,\Tq,\Tb)  = 
\nonumber && \nonumber \\ && \nonumber \hspace*{-15mm} 
 i{a^6 \over 6}\sum_{k}\sum_{\vecy,\vecz}\,
  \left\langle
  (\vimpr)_0(x)\,\zetabar_{\rm b}(\vecy)\gamma_k\zeta_\light(\vecz)
  \right\rangle  \,, 
\\
%
\hspace*{-5mm}\kvk(x_0,\Tq,\Tb)  = 
\nonumber && \nonumber \\ && \nonumber \hspace*{-15mm} 
 -{a^6 \over 6}\sum_{k}\sum_{\vecy,\vecz}\,
  \left\langle
  (\vimpr)_k(x)\,\zetabar_{\rm b}(\vecy)\gamma_k\zeta_\light(\vecz)
  \right\rangle  \,,
\\
%
\hspace*{-5mm}\kvuu
(x_0,\Tq,\Tb)  =  
\nonumber && \nonumber \\ && \nonumber \hspace*{-15mm} 
-{a^6 \over 2}\sum_{\vecy,\vecz}\,
  \left\langle
  (\vimpr)_1(x)\,\zetabar_{\rm b}(\vecy)\gamma_1\zeta_\light(\vecz)
  \right\rangle  \,,
\ees
\ei
where the label b refers to heavy {\em relativistic} quarks of mass close to the b-quark mass, and
the subscript I stands for O$(a)$ improvement, 
as discussed in~\cite{Luscher:1996sc,Sint:1997jx} for Wilson quarks.
The correlation functions are illustrated in \cite{sfb:C1a}, where 
it is also explained how they relate to quantum mechanical Hilbert space 
matrix elements. 

\begin{figure}[ht!]
\begin{center}
\vskip -6mm
\includegraphics[width=0.525\textwidth]{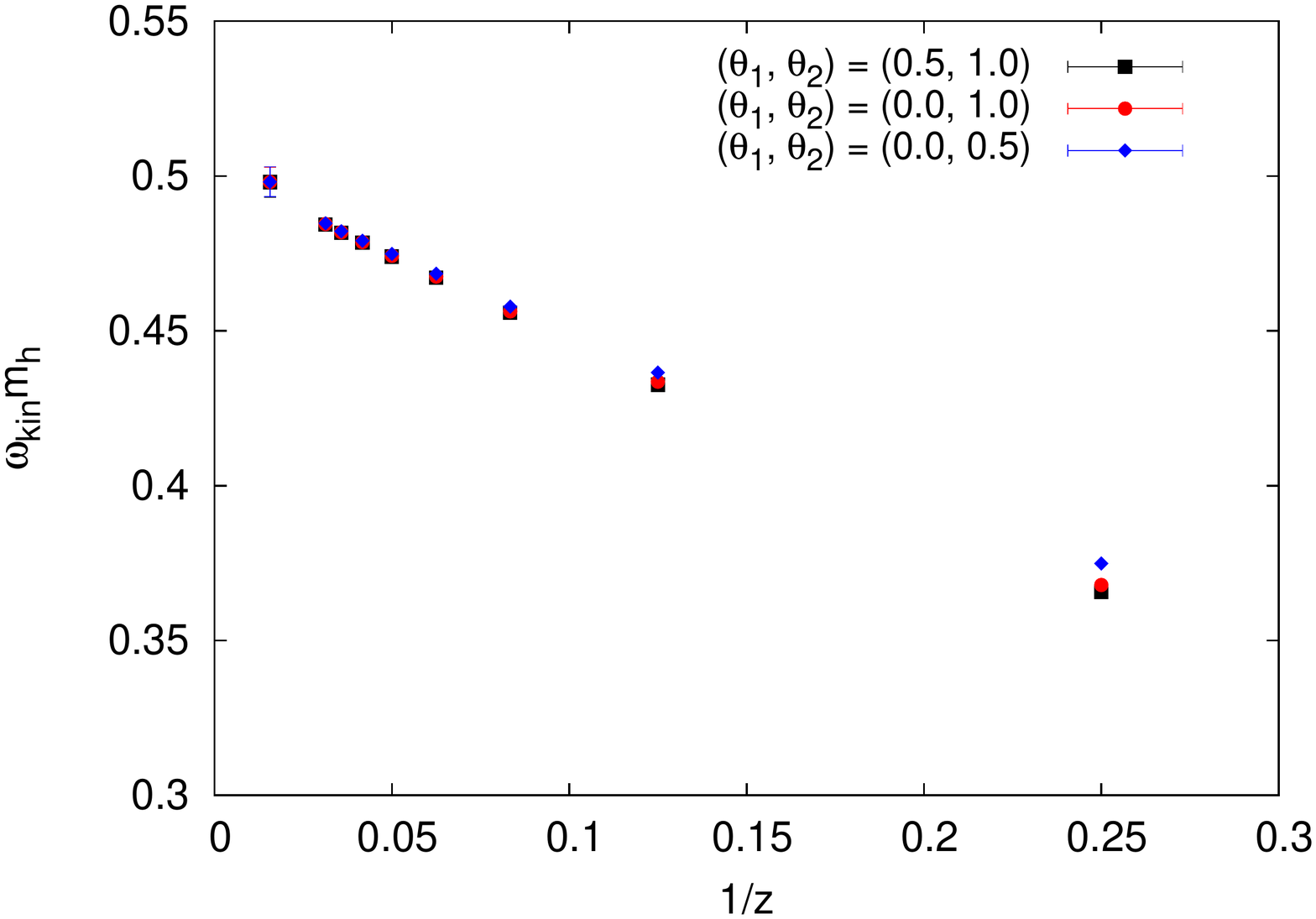} 
\vskip -6mm
\caption{%
Tree-level continuum results for the parameter $\omega_{\rm kin}$ in 
$\lag{HQET}$.
}
\label{fig:wkin}
\end{center}
\end{figure}
\begin{figure*}[t!]
\begin{center}
\vskip -6mm
\includegraphics[height=0.35\textwidth]{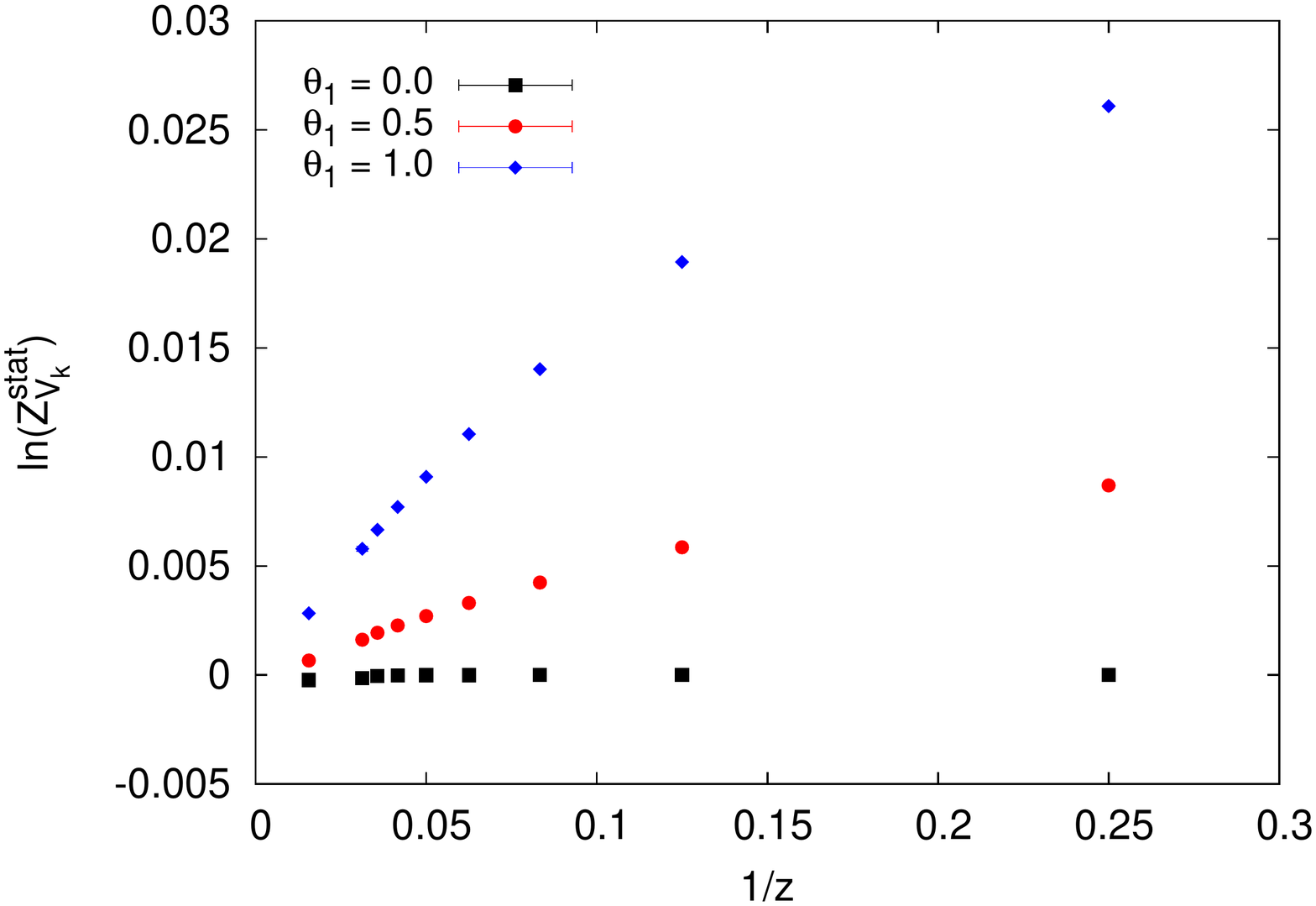} 
\includegraphics[height=0.35\textwidth]{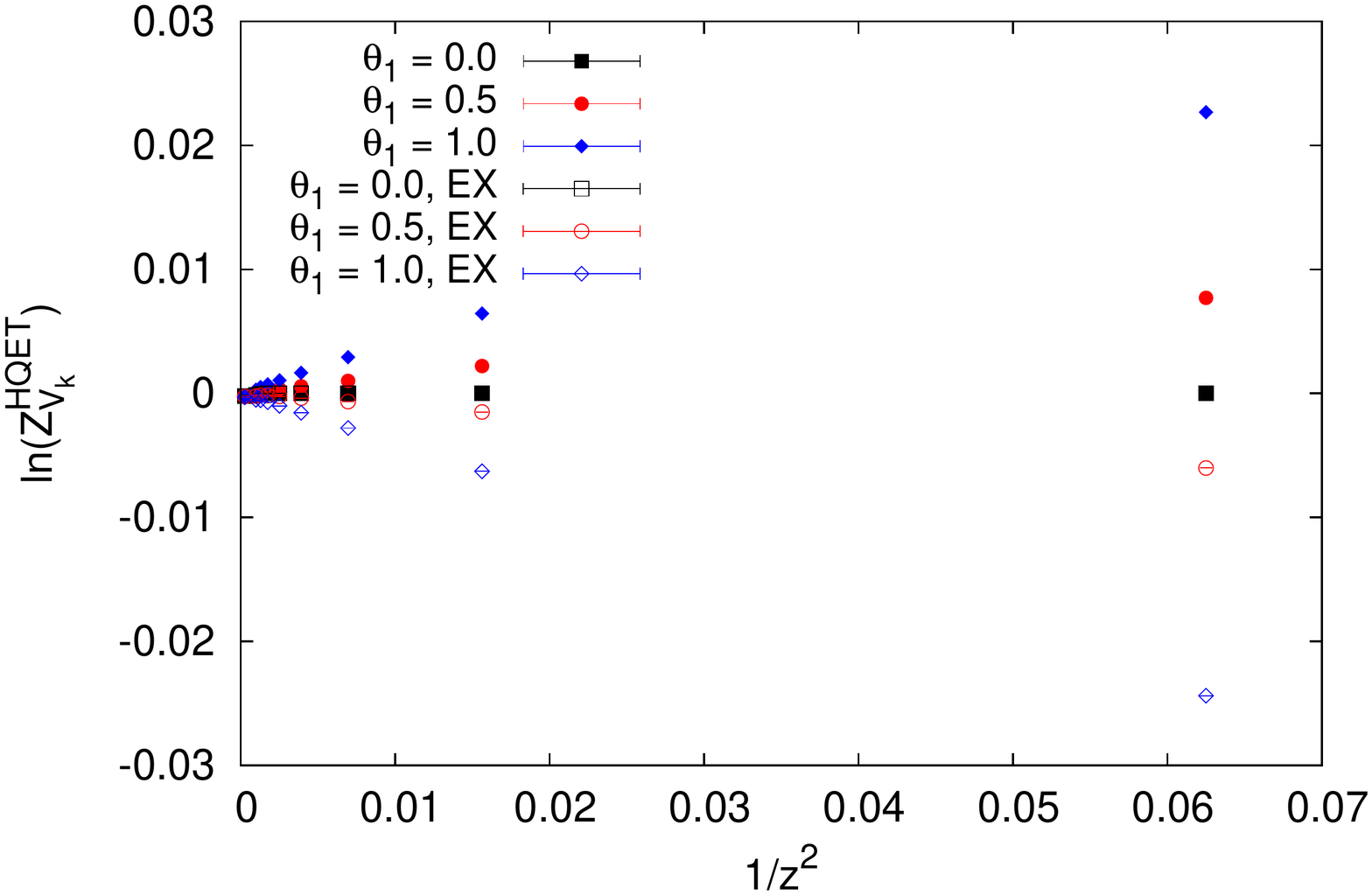} 
\vskip -6mm
\caption{%
Tree-level continuum results for the parameter $\ln Z_{\vec{\rm V}}$ of the 
spatial components of the vector current, at static order versus $1/z$ (left) 
and at $\Order{1/m_{\rm h}}$ versus $1/z^2$ (right).
$\T_1=0$ is clearly preferable, while already at $\T=(1,1,1)$ the 
corrections become very substantial.
We also note that at tree-level and for $\T_1=0$ the dependence of the 
matching equations involving a single angle on any other parameter vanishes.
Hence, the corresponding parameters are independent of whether or not using 
the "EX" setup, and higher-order contributions are absent.}
\label{fig:lnZVk}
\end{center}
\end{figure*}

In order to determine the parameters $\omega_i$, $1\leq i \leq 19$, 
the above correlation functions as well as some related
to the axial current are combined to 
19 observables~\cite{DellaMorte:2013ega,Hesse:2012hb}.
First of all, ratios are formed such that the renormalization of 
the boundary quark fields drops out. Then, one searches for combinations
and particular choices of $\T_i$ such that the matrix $\varphi$ has many 
zeros and is easily inverted.\footnote{More generally, the matrix $\varphi$
should have a good condition number.} Finally, one looks for choices where
higher $\rmO(\mhinv^2)$ terms are rather suppressed.
Such a search has successfully been carried out 
at tree-level of perturbation theory~\cite{DellaMorte:2013ega},
and we discuss some of the results in the next section.

\begin{figure}[ht!]
\begin{center}
\vskip -6mm
\includegraphics[width=0.525\textwidth]{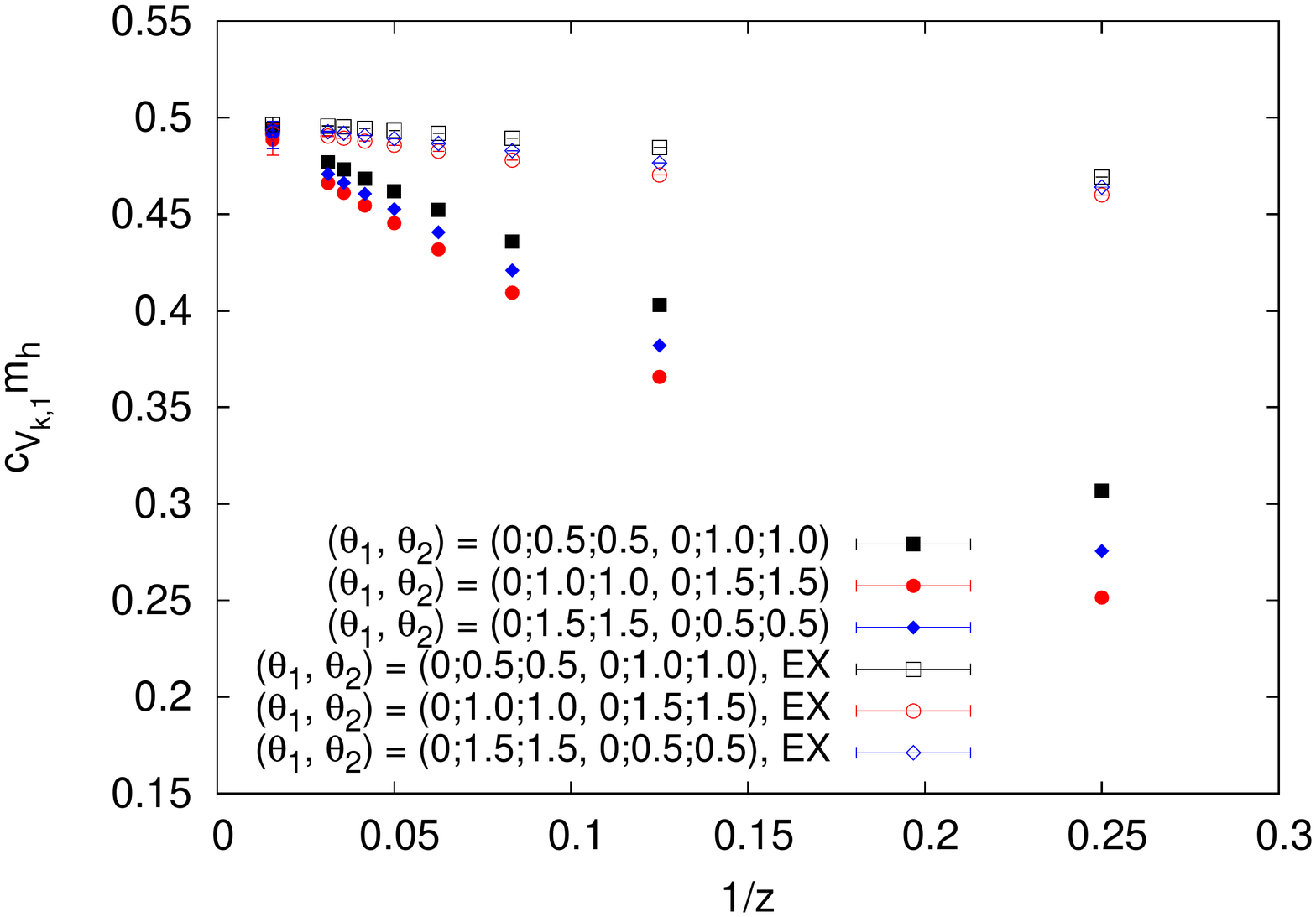} 
\vskip -6mm
\caption{%
Tree-level continuum results for the parameter $\ceff{V}{k}{1}$ (depending
on two angles) 
of the spatial components of the vector current.
Smaller values of $(\T_1,\T_2)$ tend to yield smaller 
$1/z$-corrections, and part of the higher-order $1/z$-dependence
(being smaller in the "EX" setup) is induced by other parameters.}
\label{fig:cVk13}
\end{center}
\end{figure}

\section{Tree-level study of HQET parameters}
\label{tree-level}

We now illustrate a few results of the tree-level computation \cite{DellaMorte:2013ega}
of the HQET parameters for the vector currents that are needed for the semi-leptonic 
B-decays.

After choosing a suitable set of observables $\Phi_i$, the HQET parameters are 
computed from \eq{e:matchingcond} inserting the HQET expansion \eq{e:phiexp}. 
At tree-level, each term in \eq{e:phiexp} separately has a continuum limit. 
Since we are here not interested in the cutoff effects, the limit is taken 
term by term, and we then solve the matching equations
\be
  \sum_j \varphi^j_i \cdot \omega_j = \Phi_i^{\rm QCD} -\eta_i
  \label{e:matchingsystem}
\ee
at $a=0$ for the unknown HQET parameters $\omega_j$.
Due to the block structure of $\varphi$ and because of additional
simplifications, which occur at tree-level, and with the choice of
twist angles according to the strategy of \cite{DellaMorte:2013ega},
one can determine the parameters one after the other from the matching
equations (\ref{e:matchingsystem}).

A main motivation of the tree-level study in~\cite{DellaMorte:2013ega} 
was to investigate the effect of higher-order corrections (in $\mhinv$) 
in the system of matching equations and to test the choice of observables.

At tree-level and in the static limit, the HQET parameters at $a=0$ must reproduce
the exactly known classical values. After multiplying each parameter with the 
proper factors of $\mh$, this limit is obtained for
$z\equiv \mh\cdot L\to \infty$.
The size of the higher-order contributions in $\mhinv$ can then be explored
by looking at the rate with which the HQET parameters converge to their 
known classical values. One expects in particular
\bea
\ceff{V}{k}{1}\,\mh    & = & \frac{1}{2} + \Order{1/z}\,, \\
\omegakin\,\mh    & = & \frac{1}{2} + \Order{1/z}\,, \\
\ln Z^\hqet_{V_0}      & = & 0 + \Order{1/z}\,, \\
\ln Z^\stat_{V_k}      & = & 0 + \Order{1/z}\,, \\
\ln Z^\hqet_{V_k}      & = & 0 + \Order{1/z^2}\,.
\eea
Any $\Order{1/z}$ deviations from the behaviour indicated in the 
equations above are to be understood as a sign of the 
$\mhinv$-corrections neglected in the static approximation of HQET.
Once all $\mhinv$-contributions are included in HQET and in its matching to 
QCD, as assumed here throughout, also the $\Order{1/z}$ corrections to the 
HQET parameters are fixed.
Consequently, deviations from a linear $1/z$-dependence signal higher-order 
corrections disregarded in the effective theory. They are also revealed 
by a non-vanishing dependence of the HQET parameters on the specific choices
for the $\T$-angles on which the observables used in the matching step
depend.

Within our tree-level study, where the HQET parameters can be determined
one after the other from \eq{e:matchingsystem}, one can turn off the 
effects of $\Order{\mhinv}$ corrections in the previous steps by replacing
all parameters determined in previous steps by their exact classical values. 
One then sees directly and only the effects of the new 
observables considered in the present block. This procedure is dubbed "EX".
We note that the difference between "EX" and the solution of the full
system is dominated by the $\mhinv$-effects in $\omegakin$ (\fig{fig:wkin}). 

\Fig{fig:wkin} shows that $\omega_{\rm kin}$ has 
rather linear $1/z$-corrections,
which have little dependence on $\T$. 
The coefficient of $1/z$ is around one, the standard magnitude which is 
to be expected. However, for the investigated set of $\Phi_i$,
the other HQET parameters show much smaller 
corrections, as we see below for a few examples and as
is discussed in more detail in~\cite{DellaMorte:2013ega}.

\begin{figure}[ht!]
\begin{center}
\vskip -6mm
\includegraphics[width=0.525\textwidth]{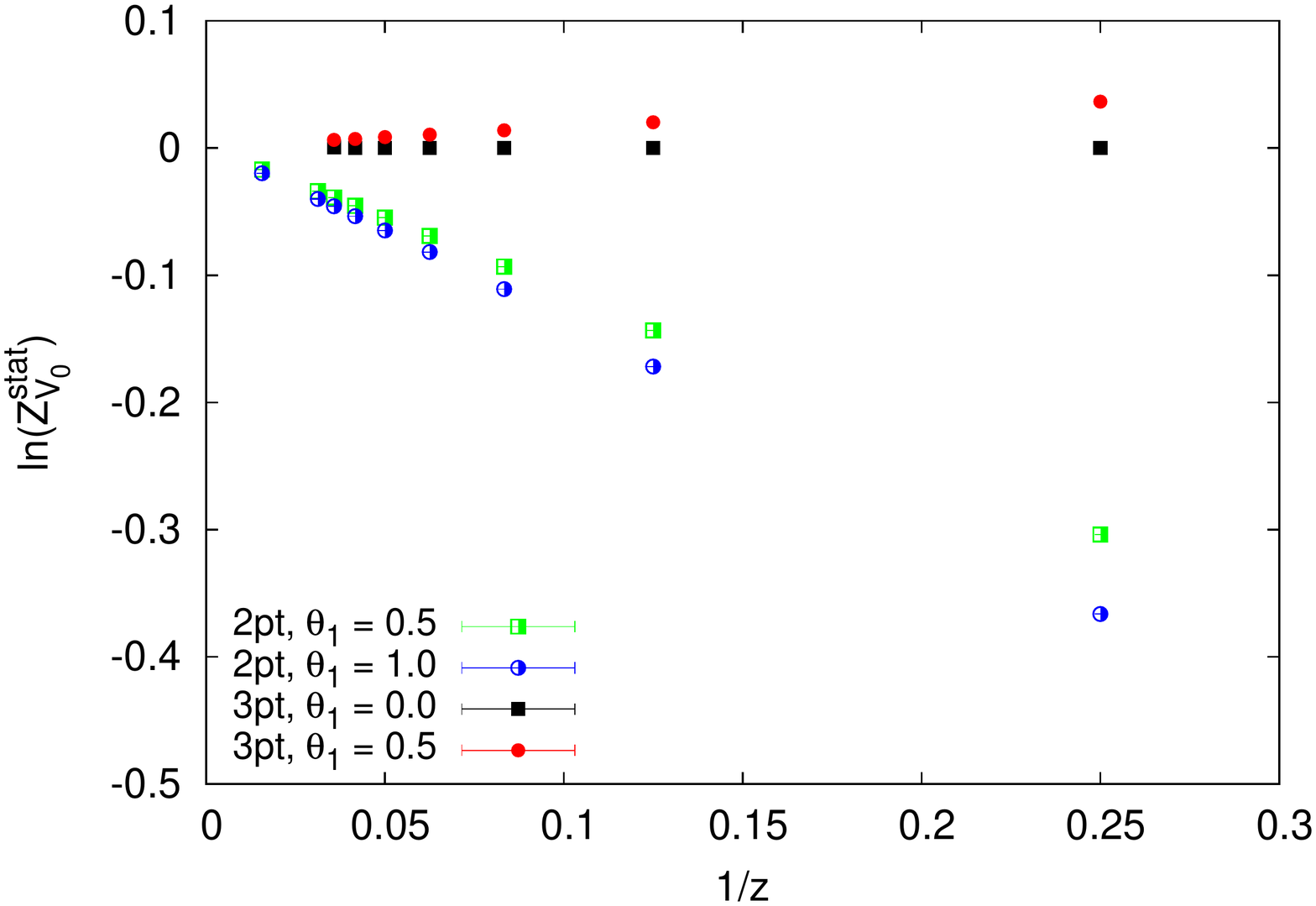} 
\vskip -6mm
\caption{%
Tree-level continuum results for the renormalization constant of the 
temporal component of the vector current, $\ln Z_{V_0}$, in the static
approximation.
We compare its definition via two-point correlation functions only (2pt)
with our standard definition to fix this HQET parameter, which uses
three-point functions (3pt).}
\label{fig:lnZV0}
\end{center}
\end{figure}
From figures \ref{fig:lnZVk}--\ref{fig:lnZV0} one can conclude that $\T_1=0$ or small is a good choice for
observables depending on a single angle (such as $\ln Z_{\vec{\rm V}}$),
whereas for the parameters via observables depending on two angles
(such as $\ceff{V}{k}{1}$) the $\T$-dependence is rather moderate.
The parameters depending on three angles (such as $\ceff{V}{k}{3}$) show a 
stronger influence by the $\T$-choices and sizable effects of higher-order 
corrections in $1/z$, which often do not decrease substantially 
(or not at all) in the "EX" case; thus, these effects appear to be a genuine 
property of them.
\Fig{fig:lnZV0} also compares two alternative ways to determine 
$\ln Z_{{\rm V}_0}^\stat$, either based on two- or three-point 
functions~\cite{DellaMorte:2013ega,Hesse:2012hb}.
Owing to the visibly flatter $1/z$-dependence, the latter is obviously to be
preferred, and $\ln Z_{{\rm V}_0}^\hqet$ and $\ln Z_{\vec{\rm A}}^\hqet$ behave
similarly.
For more details we refer again to~\cite{DellaMorte:2013ega}.
The matching strategy of \cite{DellaMorte:2013ega} 
is also studied at one-loop order in 
perturbation theory, see~\cite{Korcyl:2013ara} for a preliminary report of 
results, which support the findings of the tree-level study. 

Let us finally come back to the higher-order corrections of $\omega_{\rm kin}$,
which were found to be large compared to all others.
They propagate to a large extent into higher-order corrections of the 
HQET parameters of $A_0$ and $V_k$, which are the current components required 
for the calculation of the ${\rm B}_{({\rm s})}$-meson decay constant and form 
factors of semi-leptonic B-meson decays to light mesons.
However, as also discussed in~\cite{DellaMorte:2013ega}, it is possible to 
reduce these higher-order effects in $1/z$ by removing them completely at 
tree-level via a straightforward modification of the matching observables. 
This will be implemented together with the non-perturbative matching computation.
Furthermore, there is the option to reduce the higher-order effects in
$\omegakin$ by a generalized set of $\Phi_i$, with less zeros in $\varphi$.
While this option may reduce the impact of $\mhinv^2$-terms further, we
emphasize once more that the proposal of \cite{DellaMorte:2013ega} 
does have $\Order{1}$ coefficients as expected generically.

\section{B $\to$ K Form Factor}
\def\teq{\!\!\!=\!\!\!}

\begin{table*}[ht!]
\begin{center}
\begin{tabular}{ccccccc}
 \toprule
 id & $T/a\,\times\, (L/a)^3$ & $a$ [fm] &  $m_\pi$ [MeV] & $m_\pi L$  & \# meas. & \# target \\ 
 \midrule
 A5  & $64 \times 32^3$  &  $0.0749(8)$         &$330$  & $4.0$ & 500 & 500 \\
 F6  & $96 \times 48^3$  &  $0.0652(6)$         &$310$  & $5.0$ & 254 & 500 \\
 N6  & $96 \times 48^3$  &  $0.0483(4)$         &$340$  & $4.0$ & 220 & 500 \\
 \bottomrule
\end{tabular} 
\caption{List of ensembles used in \cite{Bahr:2014iqa}.}\label{tab:ens}
\end{center}
\end{table*}

In this section, we summarize the exploratory work of the
ALPHA Collaboration to determine
the form factors for $\mathrm B_\mathrm s \to \mathrm K \ell \nu$ decays
non-perturbatively on the lattice with $N_\mathrm f = 2$ sea quarks \cite{Bahr:2012qs,Bahr:2012vt,Bahr:2014iqa}.
For this decay no experimental data is available yet. However, the presence
of a heavier spectator s-quark renders the lattice computations technically 
simpler than for $\mathrm B \to \mathrm \pi \ell \nu$. Thus, it offers a 
good basis to gain solid control on the systematic errors.

The decay amplitude for $\mathrm B_\mathrm s \to \mathrm K \ell \nu$ is proportional 
to $|V_\mathrm{ub}|$ times the hadronic matrix element 
parameterized as
\bes
 && \hspace*{-3em} 
 \big\langle \mathrm K (p_\mathrm K) \big| V^\mu \big| \mathrm B_\mathrm s(p_{\mathrm B_\mathrm s} ) \big\rangle 
 = \nonumber\\
 & = &
 {f_+(q^2)}\, \bigg[ p_{\mathrm B_\mathrm s}^\mu + p_\mathrm K^\mu - \frac{m_{\mathrm B_\mathrm s}^2 -m_\mathrm K^2}{q^2} q^\mu \bigg] 
 \nonumber\\
 & + &
 {f_0(q^2)}\, \frac{ m_{\mathrm B_\mathrm s} ^2 - m_\mathrm K^2}{q^2} q^\mu \,,
\ees
where the two form factors $f_0(q^2)$ and $f_+(q^2)$ depend
on $q^2 = (p_{\mathrm B_\mathrm s} - p_\mathrm K)^2$, the invariant mass of the lepton pair.
When neglecting the lepton masses, only $f_+(q^2)$ contributes to the decay rate.

In order to keep the value of $q^2$ fixed in the continuum extrapolation of the form factors,
we employ flavour twisted boundary conditions, \eq{e:theta},
for the s-quark. 
By choosing the vector of twist angles, $\T$, of the s-quark, 
one can thus freely tune the momentum of the kaon. 
To remain in the rest frame of the $\mathrm B_\mathrm s$-meson, 
the heavy quark is twisted by the same angle.

We perform our measurements on gauge field ensembles generated with $N_\mathrm f =2$
dynamical sea quarks within the CLS effort. They were simulated with non-perturbatively 
O$(a)$ improved Wilson fermions and the scale is set via $f_\mathrm K$ \cite{Fritzsch:2012wq}, i.e., physical units are derived from $f_\mathrm K = 155\;\MeV$. 
All ensembles have $m_\pi L \gtrsim 4$ where finite-volume effects can be
neglected. In this work, we use the three CLS ensembles A5, F6 and N6,
described in~\cite{Fritzsch:2012wq}. 
Our error estimates take into account correlations and autocorrelations \cite{Schaefer:2010hu}. 

In the exploratory study presented here, we restrict
our analysis to the static approximation because the 
non-perturbative determination of the full set of HQET
parameters is still in progress. Thus we set
$\omega_\mathrm{kin} = \omega_\mathrm{spin} = c_{\mathrm V} = 0$.
For the renormalization constants to be used at static order
in eqs.~(\ref{e:v0hqet})~and~(\ref{e:vkhqet}),
we follow the lines of \cite{Sommer:2010ic,Heitger:2004gb}
and write
\bean
  Z^\mathrm{HQET}_\mathrm{V_0}
  &\teq& C_\mathrm{PS} (M_\mathrm h / \Lambda_{\overline{\mathrm{MS}}}) \,
  Z_\mathrm{A,RGI}^\mathrm{stat} (g_0) \,
  Z_\mathrm{V/A}^\mathrm{stat} (g_0) \, , \label{Zv0renorm} \\
  Z^\mathrm{HQET}_\mathrm{V_k} &\teq& C_\mathrm{V}  (M_\mathrm h / \Lambda_{\overline{\mathrm{MS}}}) \,
  Z_\mathrm{A,RGI}^\mathrm{stat} (g_0)\, . \label{Zvkrenorm}
\eean
The matching factors, $C_\mathrm{PS}$ and $C_\mathrm{V}$, 
are known at three loops in perturbation theory \cite{Chetyrkin:2003vi},
and $Z_\mathrm{A,RGI}^\mathrm{stat}$ is known non-perturbatively \cite{DellaMorte:2006sv}.
The truncation of the theory at static order is expected to be a 10-20\% effect.

For an analysis in the static approximation, we only need to consider the two- 
and three-point correlation functions (while in the same measurement runs we 
also determine the various correlation functions that are needed for the 
extraction of the form factors in HQET at order $1/m_\mathrm h$)
\bean
 \mathcal C^{\mathrm K} (x^0-y^0; \vecp)  & \teq  & \sum\limits_{\vecx, \vecy} 
 \mathrm e^{-\mathrm i \vecp \cdot (\vecx - \vecy)} 
 \langle P^{\mathrm s \mathrm u} (x) P^{\mathrm u\mathrm s }(y)  \rangle
 \,, \\
 \mathcal C^{\mathrm B}_{ij} (x^0-y^0; \mathbf{0})  & \teq  & \sum\limits_{\vecx, \vecy} 
 \langle P^{\mathrm s\mathrm b}_i(x) P^{\mathrm b\mathrm s}_j(y)  \rangle
 \,, \\
 \mathcal C^{3}_{\mu,\,j}(t_\mathrm K, t_\mathrm B; \vecp) & \teq & \sum\limits_{\vecx_{\mathrm K}, \vecx_V, \vecx_{\mathrm B}} 
 \mathrm e^{-\mathrm i \vecp \cdot (\vecx_{\mathrm K} - \vecx _V)} 
 \nonumber\\
 && \times\ \langle P^{\mathrm s\mathrm u}(x_{\mathrm K}) {\mathrm V}_\mu^\stat(x_V) P^{\mathrm b\mathrm s}_j (x_{\mathrm B}) \rangle \, . 
\eean
The $P^{\mathrm{q}_1 \mathrm{q}_2}_i(x)$ are interpolating fields, such as 
$\overline{\psi}_{\mathrm{q}_1} (x) \gamma_5 \psi_{\mathrm{q}_2} (x)$, for the mesons. 
Different levels of Gaussian smearing \cite{Gusken:1989ad} 
of the s-quark in the heavy-light meson, i.e., different trial wave functions,
are indicated by the subscripts $i$ or $j$. For any suitable smearing $i$, the ratio
\begin{equation}
 f_{\mu,\,i}^\mathrm{ratio} (t_\mathrm B, t_\mathrm K; q^2) \equiv
 {\frac{\mathcal C^{3}_{\mu,\,i} (t_\mathrm K, t_\mathrm B)}
 {\sqrt{ \mathcal C ^\mathrm K (t_\mathrm K) \mathcal C^{\mathrm B}_{ii} (t_\mathrm B) }} \,  
 \mathrm e^{\frac{E_\mathrm K t_\mathrm K}{2}}\mathrm e^{\frac{E_\mathrm B t_\mathrm B}{2}}}
\label{fratio}
\end{equation}
will give the desired matrix element in the limit of large Euclidean times, 
$t_\mathrm B \equiv x^0_{\mathrm B} - x^0_V$ and $t_\mathrm K \equiv x^0_V - x^0_{\mathrm K}$:
\begin{equation}
 \langle \mathrm K(p_\mathrm K^{\theta}) | {\mathrm V}_\mu | \mathrm B_\mathrm s(0 ) \rangle = 
  \lim_{t_\mathrm B, t_\mathrm K \to \infty}  f_{\mu,\,i}^\mathrm{ratio} (t_\mathrm B, t_\mathrm K; q^2)\,.
 \label{fme} 
\end{equation}
Alternatively, we can parameterize the $t_\mathrm B$- and $t_\mathrm K$-dependence of the 
correlation functions as
\bean
 \mathcal C^{\mathrm K} (t_\mathrm K)  & \teq & \sum _m (\kappa^{(m)})^2 \mathrm e^{-E^{(m)}_\mathrm K t_\mathrm K}
 \,, \\\
 \mathcal C_{{ij}}^{\mathrm B} (t_\mathrm B)  & \teq & \sum _n \beta_{i}^{(n)} \beta_{j}^{(n)} \mathrm e^{-E^{(n)}_\mathrm B t_\mathrm B}
 \,, \\
 \mathcal C_{\mu,\,i}^3 (t_\mathrm B, t_\mathrm K) & \teq & \sum _{m,n} \kappa^{(m)} {\varphi_\mu^{(m,n)}} \beta_{i}^{(n)} 
 \mathrm e^{-E^{(n)}_\mathrm B t_\mathrm B} \mathrm e^{-E^{(m)}_\mathrm K t_\mathrm K}\,,
\eean
and determine $\{\kappa^{(m)}, E^{(m)}_\mathrm K\}$ from a fit to $\mathcal C^{\mathrm K}$,
and $\{ \beta_{i}^{(n)}, \varphi_\mu^{(n,m)}, E^{(n)}_\mathrm B\}$ from a combined
fit
to $\mathcal C_{\mu,\,{i}}^3$ and $\mathcal C_{{ij}}^{\mathrm B}$.
The matrix element of eq.~(\ref{fme}) is then given by the
fit parameter $\varphi_\mu^{(1,1)}$.
We include the first excited $\mathrm B_\mathrm s$-state, but only the kaon ground state
(i.e., we take $m=1$ and $n=1,2$).

The ratio $f^\mathrm{ratio}_\mu$ of eq.~(\ref{fratio})
is shown as a function of $t_\mathrm B$ at fixed $t_\mathrm K=20$ 
in figure~\ref{fig:N6}. For comparison, the value of $\varphi^{(1,1)}_\mu$ 
resulting from the fit is indicated as a red band.

\begin{figure}[ht!]
\begin{center}
\scalebox{0.45}{\includegraphics{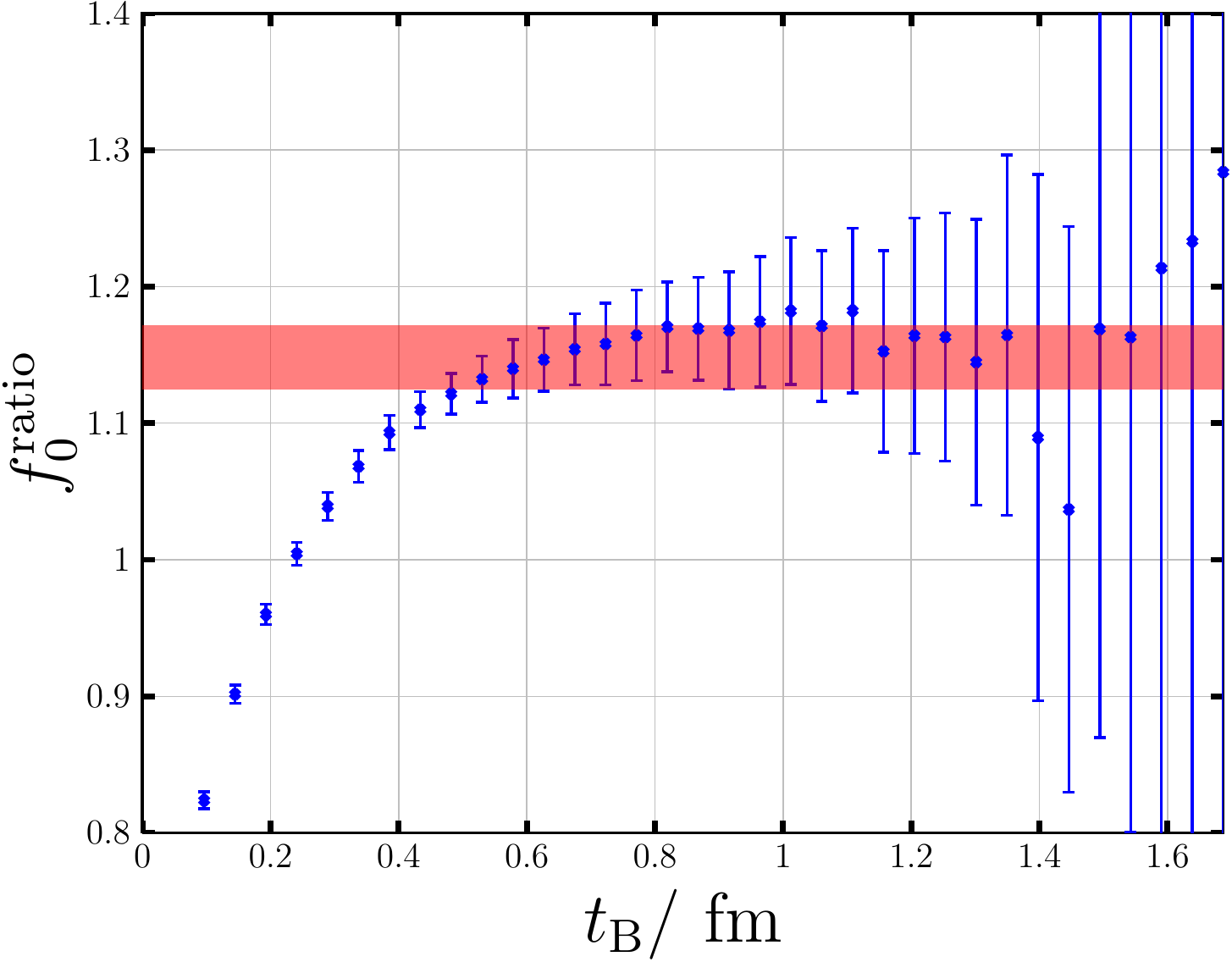}}
\caption{The ratio $f^\mathrm{ratio}_{\mu}$ (blue points) and the fit result $\varphi_\mu^{(1,1)}$ (red band) 
for lattice N6, $\mu=0$ and fixed $t_\mathrm K = 20$.
} 
\label{fig:N6}
\end{center}
\end{figure}
In the rest frame of the $\mathrm B_\mathrm s$-meson, the matrix elements have the form
\bea
  \langle \mathrm K | {\mathrm V}_0 |\mathrm B_\mathrm s\rangle 
  & = &\sqrt{2 m_{\mathrm B_\mathrm s}}\, f_\parallel(q^2)\,, \qquad
  \nonumber\\
  \langle \mathrm K | {\mathrm V}_i |\mathrm B_\mathrm s\rangle 
  & = & \sqrt{2 m_{\mathrm B_\mathrm s}}\, p_\mathrm K ^i\, f_\perp(q^2) \,,
  \nonumber
\eea
where the form factors $(f_\parallel, f_\perp)$ are related to $(f_+,f_0)$ by
\bes
  f_+ & \teq & \frac{1}{ \sqrt{2 m_{\mathrm B_\mathrm s}}}\, f_\parallel  
     +\frac{1}{\sqrt{2m_{\mathrm B_\mathrm s}}}(m_{\mathrm B_\mathrm s}-E_\mathrm K)\, f_\perp \,, \label{fpl} 
  \\
  f_0 & \teq & \frac{ \sqrt{2 m_{\mathrm B_\mathrm s}}}{ m_{\mathrm B_\mathrm s}^2-m_\mathrm K^2} 
  \Bigl[(m_{\mathrm B_\mathrm s}-E_\mathrm K)f_\parallel\nonumber\\
  & & +\,(E_\mathrm K^2-m_\mathrm K^2)f_\perp \Bigr] \,. 
\ees

The form factor $f_+$ extracted from the fitted $\varphi^{(1,1)}_\mu$
at different lattice spacings is shown in figure~\ref{fig:cont}.
When computing $f_+$ in the static approximation of HQET,
we are free to keep or drop terms of order $1/m_\mathrm h$ in eq.~(\ref{fpl}).
To illustrate the corresponding $\Order{1/m_\mathrm h}$ ambiguity, we show 
in figure~\ref{fig:cont} (and \ref{fig:comp}) two sets of data points:
the upper one corresponds to using all terms in eq.~(\ref{fpl}),
the lower one to dropping the term proportional to $f_\parallel$.
This ambiguity will disappear, once all $\Order{1/m_\mathrm h}$ terms of HQET are included.
For both sets we show a constant continuum extrapolation and one linear in $a^2$.
The latter has by far the larger error, and within this error it is consistent with the 
result of the constant extrapolation.

\begin{figure}[t!]
  \scalebox{0.450}{\includegraphics{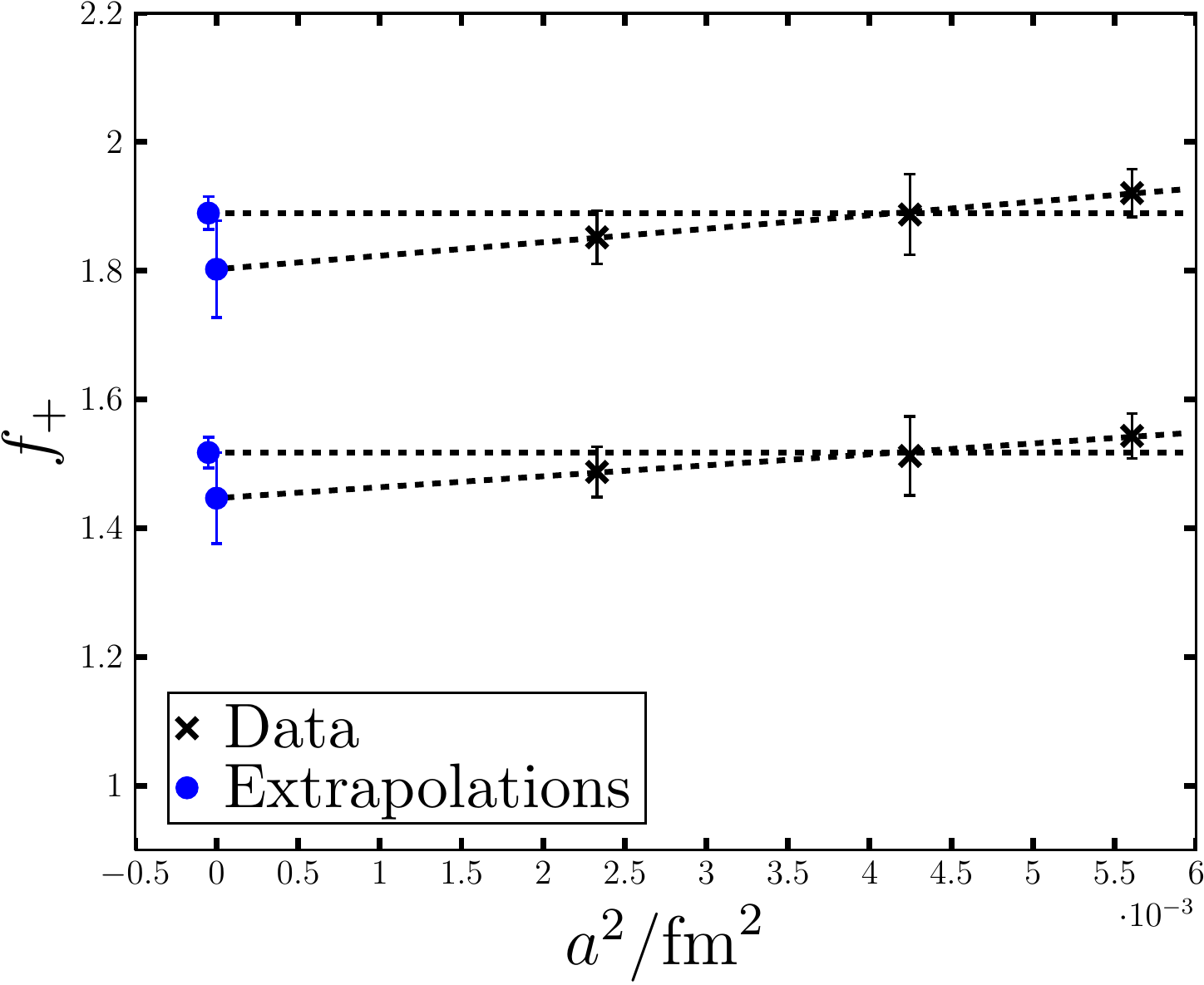}}
  \caption{Continuum extrapolation of our data for $f_+$ at $q^2 = 21.23\,\mathrm{GeV}^2$.}
  \label{fig:cont}
\end{figure}

Figure~\ref{fig:comp} shows a comparison of our results from the linear continuum extrapolation
of $f_+(q^2)$ with recent results of HPQCD \cite{Bouchard:2014ypa} (at their smallest 
$a = 0.09\,\mathrm{fm}$ and $m_\pi = 320\, \mathrm{MeV}$).

\begin{figure}[t!]
  \scalebox{0.45}{\includegraphics{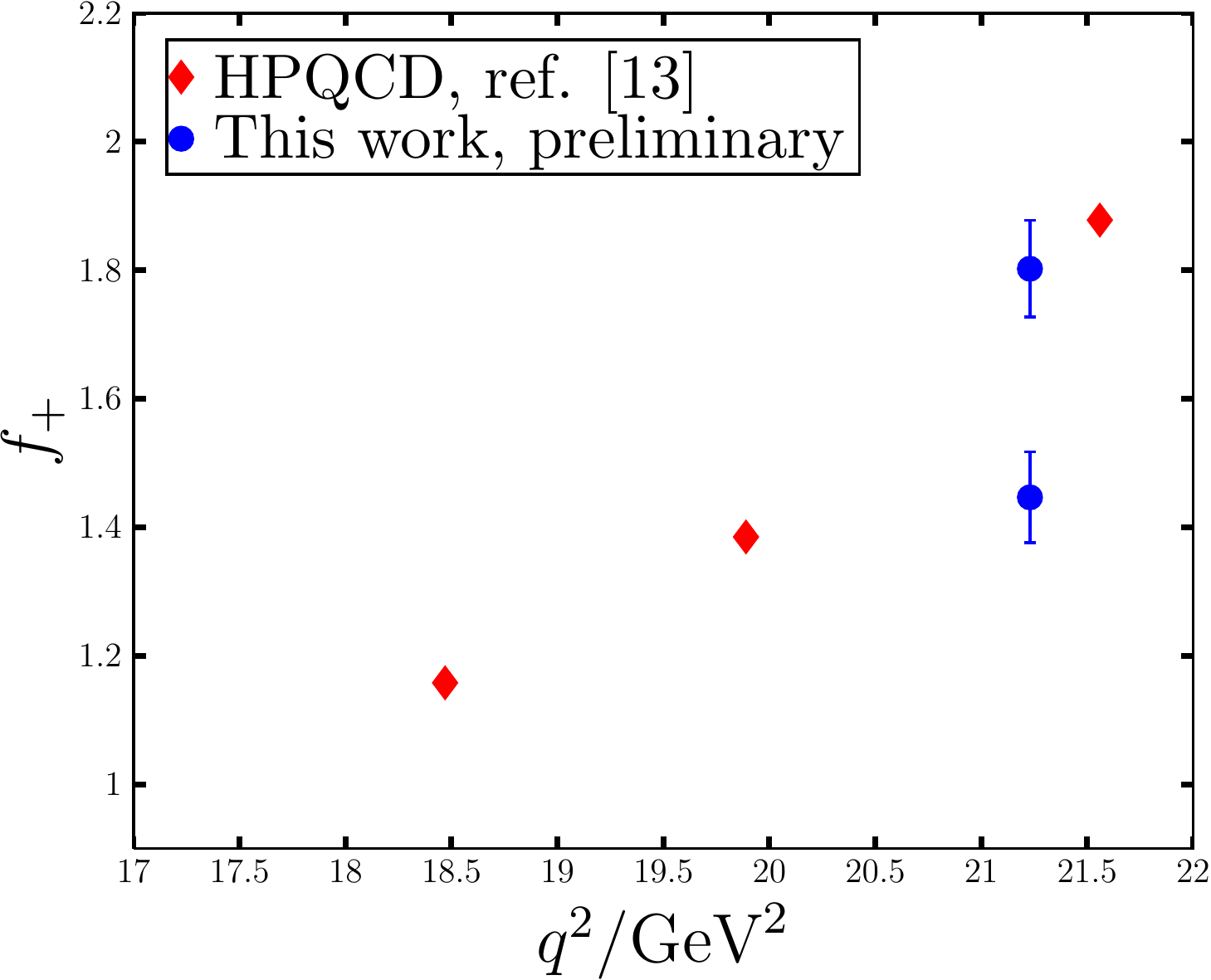}}
  \caption{Comparison of lattice QCD results at various values of $q^2$.}
  \label{fig:comp}
\end{figure}

\section{Conclusions}
\label{concl}

With the matching conditions between HQET at $\Order{1/m_{\rm h}}$ and QCD
reviewed here, it has been demonstrated that the
solution of the extended matching problem, including all components of 
the axial and vector currents, is feasible.
Our lattice formulation basically amounts to solving a linear system of 
matching equations with the 19 HQET parameters as unknowns.
The matrix associated with this system has a simple block structure.
The realization of this matching strategy employs observables
which are constructed from correlation functions in a finite volume with 
Schr\"odinger functional boundary conditions, and which would correspond
to mesonic correlation functions in large volume. We exploit particular 
kinematic settings that are well suited for a straightforward 
implementation on the non-perturbative level by numerical simulations.

Although our matching strategy (entirely expressed through renormalized 
quantities in QCD) is not necessarily restricted to Wilson fermions,
we will use these in the future non-perturbative computation.
According to our past experience with Wilson fermions~\cite{Blossier:2010jk,Blossier:2012qu},
very fine lattice spacings and good numerical precision could be achieved.
Moreover, with our tree-level study of the $1/z$-dependence of the HQET 
parameters we have illustrated that the linear behaviour already sets in for 
$z\geq 8$ in most cases, with slopes of $\Order{1}$. 
In fact, this finding is very reassuring, as it implies that for the 
non-perturbative framework followed by our 
collaboration~\cite{Blossier:2012qu}, with a matching volume of extent
$L \approx 0.5$ fm and dimensionless heavy quark masses around $z=13$ at 
the b-quark mass scale, higher-order corrections beyond the already included
$\Order{1/m_{\rm h}}$ ones are suppressed by a factor of about 10.

The non-perturbative solution of the HQET matching problem is a crucial 
step towards a precision computation of phenomenologically relevant 
hadronic matrix elements in B-physics. We presented here 
the current status of our effort to compute 
form factors for semi-leptonic B-decays using HQET on the lattice.  
In particular, we considered the decay $\mathrm B _\mathrm s \to \mathrm K \ell\nu$ 
and investigated two different methods to extract the form factor $f_+(q^2)$ 
at a fixed value of $q^2 = 21.23\,\mathrm{GeV}^2$ from 
the plateau value of a suitable ratio of correlators or from 
a simultaneous fit to the functional form of the correlators.

The authors of~\cite{Bahr:2014iqa} performed a continuum extrapolation of the lattice data 
for $f_+(q^2)$ and found very small O$(a^2)$ effects. While these preliminary 
results were still computed in the static approximation and an extrapolation 
to the physical pion mass has yet to be performed, the preliminary value 
of $f_+$ is in agreement with the results from other collaborations
at the same quark mass.

There is thus no sign of reduction of the difference of $V_\mathrm{ub}$
extracted from semi-leptonic and inclusive/leptonic decays, but at the moment
there are 10-20\% uncertainties due to neglecting the $1/m_{\rm h}$-terms.
These terms have to be included for a firm conclusion.
Once the HQET parameters are known non-perturbatively, 
they will all be included in the analysis. Moreover, we plan to extend 
the computation to $\mathrm B\to\pi\ell\nu$ decays, several values of $q^2$, 
and $N_\mathrm f=2+1$ flavours of sea quarks.

\vskip2.5em
\noindent
{\bf Acknowledgements.}
We thank Felix Bahr, Fabio Bernardoni, John Bulava, Michael Donnellan,
Samantha Dooling, Dirk Hesse, Anosh Joseph and Piotr Korcyl
for a pleasant collaboration and many useful discussions.
In particular we received many valuable comments by Piotr Korcyl
on the topics of this review.
\\
We are grateful for the support of the 
Deutsche Forschungsgemeinschaft (DFG)
in the SFB/TR~09 ``Computational Particle Physics'' and 
we have profited very much from the scientific exchange  in the SFB.
This work was partially supported by the Spanish Minister of Education and 
Science, project RyC-2011-08557 (M.~D.~M.), and by the grant HE~4517/3-1
of the DFG (J.~H.). 
\\
We gratefully acknowledge the Gauss Centre for Supercomputing (GCS)
for providing computing time through the John von Neumann Institute for
Computing (NIC) on the GCS share of the supercomputer JUQUEEN at J\"ulich
Supercomputing Centre (JSC). GCS is the alliance of the three national
supercomputing centres HLRS (Universit\"at Stuttgart), JSC (Forschungszentrum
J\"ulich), and LRZ (Bayerische Akademie der Wissenschaften), funded by the
German Federal Ministry of Education and Research (BMBF) and the German State
Ministries for Research of Baden-W\"urttemberg (MWK), Bayern (StMWFK) and
Nordrhein-Westfalen (MIWF).
We acknowledge PRACE for awarding us access to resource JUQUEEN in Germany at J\"ulich and to resource SuperMUC based in Germany at the LRZ, Munich. 
We also thank the LRZ for a CPU time grant on SuperMUC, project pr85ju, and
DESY for access to the PAX cluster in Zeuthen.





\end{document}